# Geometrical destruction of the global phase coherence in ultrathin superconducting cylinders

Y. Liu, Yu. Zadorozhny, M.M. Rosario, B.Y. Rock, P.T. Carrigan, H. Wang

*Department of Physics, The Pennsylvania State University, University Park, PA 16802, U.S.A.*

**The global phase coherence in doubly-connected superconductors leads to fluxoid quantization, allowing the superfluid velocity $v_s$ to be controlled by an applied magnetic flux. In ultrasmall samples this quantization requirement leads, surprisingly, to the destruction of the phase coherence itself around half-integer flux quanta, because of the sample-size-induced growth in $v_s$, as predicted by de Gennes. We report observations of the predicted phenomenon in ultrathin Al and $Au_{0.7}In_{0.3}$ cylinders, and the corresponding phase diagram for ultrathin superconducting cylinders. The new phase diagram features disconnected superconducting regions, as opposed to the single one seen in the conventional Little-Parks experiment.**

Recent advances in nanoscience have demonstrated that fundamentally new physical phenomena may be found when the size of samples shrinks. In the area of superconductivity, the reduction of sample size has led to the observation of the paramagnetic Meissner effect in micron-size superconductors (*1*), the quantization of the Bose condensate in submicron samples (*2*), and ultimately the suppression of superconductivity in nanometer-scale superconductors (*3,4*). In this regime, it has also been recognized that the sample topology has particularly strong effects on superconductivity, as reflected in the characteristic features of the phase diagrams for singly- and doubly-connected samples (*5,6*).

A unique feature of doubly-connected superconductors (independent of the sample size) is fluxoid ($\Phi'$) quantization (7,8) in the units of $\Phi_0 = h/2e$, where h is the Planck constant, and e is the electron charge, due to the presence of global phase coherence among the Cooper pairs (9). The fluxoid $\Phi'$ is defined by

$$\Phi' = \Phi + \left(\frac{m^*c}{e^*}\right)\oint_C \mathbf{v}_s \, ds, \qquad (1)$$

where $\Phi (= \int H \, dS = \oint_C A \, ds)$ is the ordinary magnetic flux, $m^*$ is the effective mass of the Cooper pairs, $v_s$ is the tangential superfluid velocity, and $C$ is a closed contour in the superconductor. If $C$ is deep in a (bulk) superconductor, $v_s$ vanishes so that $\Phi' = \Phi$. For a cylindrical film with a thickness smaller than the penetration depth, $v_s$ is uniform in the sample (10). Because of fluxoid quantization, for a given flux $\Phi$,

$$v_s = (2\hbar/m^*d)(n-\Phi/\Phi_0) \qquad (2)$$

where $d$ is the cylinder diameter, and $n$ is an integer that minimizes $v_s$. This leads to the well known Little-Parks effect (11), characterized by a small oscillation in $v_s$ that results in an oscillation in the superconducting transition temperature ($T_c$) and the sample resistance in the transition regime, with a period of $\Phi_0$.

A dramatic consequence of fluxoid quantization in ultrasmall superconductors was pointed out by de Gennes (5). He considered, within the phenomenological Ginzburg-Landua theory, the behavior of a superconducting ring with a side arm of length $L$ as its diameter $d$ was varied. For the special case of $L = 0$, a simple ring geometry was recovered. Two very different physical regimes were predicted for different ring diameters. For large rings, the conventional Little-Parks effect was found, and superconductivity was shown to exist at zero-temperature in all magnetic fields up to the critical field. However, for $d < \xi(0)$, where $\xi(0)$ is the zero-temperature superconducting coherence length, a destructive regime was predicted. Away from half-integer flux quanta, the phase boundary was found to be the same as that for the larger rings. However, for

$$(k\Phi_0 - \delta)/2 < \Phi < (k\Phi_0 + \delta)/2 \qquad (3)$$

where $k$ is an odd integer and $\delta = (1 - d/\xi(0))\Phi_0$, it was demonstrated that superconductivity was not possible, even at zero-temperature.

The predicted destruction of superconductivity at zero-temperature is intimately related to sample geometry. Within the Ginzburg-Landau free energy, the kinetic energy density of the supercurrent, $\frac{1}{2} n_s^* m^* v_s^2$, where $n_s^*$ is the number density of the Cooper pairs, can be compared with the superconducting condensation energy density, $H_c^2/8\pi = n_s^* \hbar^2/4m^* \xi^2(T)$, where $H_c$ is the thermodynamic critical field. Equation 2 suggests that the doubly-connected sample geometry demands that $v_s$ increase towards its maximum value of $v_s^{max} = \hbar/m^* d$ at half-integer flux quanta, as long as the global phase coherence is present in the sample. If $d$ is made sufficiently small, the kinetic energy would be pushed so high (as the flux increases) that it would be impossible to compensate this energy by the condensation energy, making the globally phase coherent superconducting state energetically unfavorable. This particular way of suppressing superconductivity is fundamentally different from that by strong disorder or Coulomb repulsion (*12*). Therefore it may be appropriate to call this phenomenon the geometrical destruction of global phase coherence.

Experimentally, this phenomenon is difficult to observe. The condition $d < \xi(0)$ requires rings of extremely small diameter and, therefore, wire thickness. These types of samples typically have short coherence lengths, because of the unavoidable disorder introduced by structural defects and boundary roughness. For example, $\xi(0)$ was found to be only 0.1 - 0.2 µm in mesoscopic Al disks, squares, and loops (*2,6*). In comparison, $\xi(0)$ should be 1.6µm in single crystalline Al (*13*). In Ref. 6, the effect of sample geometry on mesoscopic superconductors was experimentally studied. Indeed, the phase diagram for a singly connected sample was found to be significantly different from that of a doubly connected loop of the same size because of the absence of orbital (vortex) states in the latter type of samples. However, the size of the samples in this previous study (1 μm, an order of magnitude larger than $\xi(0) = 0.1$ μm) were too large to reach the regime considered theoretically by de Gennes.

Ultrathin cylinders have advantages over the ring for detecting the destructive regime since its parallel critical field, $H_{c//}(T)$, can be high, while at the same time the superconducting coherence length can be reasonably long. Therefore, ultrathin superconducting cylinders, rather than rings prepared by conventional nanolithographic techniques, were chosen for the present study. Here we report the observation of the destructive regime in ultrathin, doubly-connected cylinders of Al and $Au_{0.7}In_{0.3}$, where the global phase coherence was indeed suppressed around half-integer flux quanta, apparently even in the zero-temperature limit.

The cylindrical samples were prepared by depositing Al or $Au_{0.7}In_{0.3}$ onto an insulating quartz filament, as previously described (*14,15*). The cylinders were ~1mm long and as small as 150 nm in diameter, nearly an order of magnitude smaller than previously studied (*7*). Electrical transport measurements were carried out in a dilution or a $^3$He refrigerator equipped with a superconducting magnet, with base temperatures of 20 mK and 0.3K respectively. The cylinders were manually aligned to be parallel to the magnetic field. The cylinder diameters were inferred from the resistance oscillation period and confirmed by atomic force microscope (AFM) measurements.

In Fig. 1, the resistance of an Al cylindrical film (Al-1, $d = 150$ nm) is plotted as a function of $\Phi$ and $T$. At low $T$, the sample was superconducting for a substantial range of magnetic field below $H_{c//}$. However, the zero sample resistance was suppressed around $\Phi = \pm\frac{1}{2}\Phi_0$, and $\pm\frac{3}{2}\Phi_0$, resulting in narrow resistance peaks. At the lowest temperature, $T = 20$ mK, the resistance peaks at $\Phi = \pm\frac{1}{2}\Phi_0$ had a magnitude $R \sim 310 \Omega$, a significant fraction of the normal-state resistance $R_N$ 930 , and a width of approximately 0.18 $_0$, as measured at the onset of nonzero resistance.

Figure 2 shows the temperature dependence of the sample resistance measured in zero and finite fields corresponding to integer and half-integer flux quanta. In zero field, Al-1 became superconducting around 1.3 K. At $\frac{1}{2}$ $_0$ its resistance showed a broad drop starting around 1K, in strong contrast with R(T) at $\Phi =$ $_0$ where a sharp transition to zero resistance was seen at 1K even though the applied field was higher. Similar behavior was also observed in an ultrathin cylinder of $Au_{0.7}In_{0.3}$ (AuIn-1, $d = 154$ nm), shown in Fig. 3. For both Al-1 and AuIn-1, R(T) at $\frac{1}{2}$ $_0$ leveled off to a substantial fraction of $R_N$, showing almost no change from 200mK down to 20 mK. In contrast, the temperature dependence of a larger Al cylinder (Al-2, $d = 357$nm), shown in Fig. 4, displayed a conventional $T_c$ oscillation with no essential difference in the shape of R(T) at integer and half-integer flux quanta.

The systematic behavior observed in all samples discussed above strongly suggests that a sample with a sufficiently small diameter may remain non-superconducting around half-integer flux quanta even at zero-temperature. A generic phase diagram can thus be obtained for ultrasmall, doubly-connected superconducting samples, as shown in Fig. 5. In this phase diagram, a normal phase extends deep into the region where superconductivity would be expected for cylinders of a conventional size. For these samples, the well-established phase diagram consists of a single superconducting region with a slightly modulated phase boundary extending up to the parallel critical field, $H_{c//}$ (Inset b of Fig. 4). In comparison, the new phase diagram is qualitatively different, featuring disconnected superconducting regions separated by a normal phase.

To quantitatively compare our experimental results with the theory, it is necessary to determine (0). Finite-temperature (T) can be estimated from $H_{c//}(T) = \sqrt{3}$ $_0/$ t (T), where t is the film thickness (*16*). Using the onset $H_{c//}(T)$, values of $\xi(T)$ are found to be 161 nm for Al-1 (*d* = 150 nm) at 20 mK, 160 nm for AuIn-1 (*d* = 154 nm) at 20 mK, and 60 nm for Al-2 (*d* = 357 nm) at 0.39K ($\xi(0) <$ 60nm). Therefore, we may conclude that $d < \xi(0)$ for both Al-1 and AuIn-1, while $d > \xi(0)$ for Al-2 (which is more disordered than Al-1), in full agreement with theoretical expectation.

Several questions of fundamental interest are raised by the results obtained in this study. A substantial drop in R(*T*) taken at half-integer flux quanta was found in Al-1 and, to a lesser extent, in AuIn-1 as well. The origin of the resistance drop is clearly related to superconductivity. As discussed above, the zero-temperature normal state observed in the destructive regime originates from the loss of global phase coherence in samples with a doubly-connected geometry. Therefore, it might be reasonable to ask whether it is possible that the local pair formation may survive, leading to a novel metallic phase of Cooper pairs. Similar states have been discussed in the context of disordered superconductors coupled with a dissipative bath (*17*). More experiments are needed to clarify the nature of the normal state found in the present experiment.

What would happen if the diameter of the cylinder were to be made even smaller? In particular, what should we expect when the circumference of the cylinder becomes smaller than the superconducting coherence length? In this limit, a Ginzburg-Landau equation in a coordinate along the circumference of the cylinder, as used in Ref. 5, is presumably invalid. A microscopic theory has not been attempted. Experimentally, the preparation of doubly connected superconducting samples of dimensions on the nanometer scale challenges the existing technologies. In this regard, superconducting carbon nanotubes (*18*) are a promising candidate for such studies.

Singly connected superconducting Al disks, in which global phase coherence was observed directly in samples of size smaller than $\xi(0)$ (*2*), have been studied experimentally (*1,2,6*) and theoretically (*19,20*), However, singly connected superconducting wires in a parallel magnetic field are yet to be explored. Novel phenomena, in particular, those associated with vortex states, may be expected (*19,20*). More surprises from these nanoscale superconductors, with or without a doubly connected sample geometry, may await.

**Fig. 1** Resistance as a function of $\Phi$ and $T$ for Al-1, an Al cylinder with diameter $d$ = 150 nm and wall thickness $t$ = 30 nm. Even at temperatures much lower than the zero-field $T_c$ (= 1.30 K at onset), the sample remained normal around $\Phi = \pm\frac{1}{2}\Phi_0$ and $\pm\frac{3}{2}\Phi_0$. At $T$ = 20 mK the resistance peak at $\Phi = \pm\frac{1}{2}\Phi_0$ has a width of $\Delta\Phi = 0.18\Phi_0$, and a magnitude of R = 0.33$R_N$, where $R_N$ is the normal state resistance. The superconducting coherence length $\xi$(20mK) is approximately 161 nm, as estimated from the parallel critical field $H_{c//}$(20mK) = 2365 G ($\Phi_{c//}$ = 2.03 $\Phi_0$). Values of resistance were taken every 0.01 $\Phi_0$ from -2.5 $\Phi_0$ to +2.5 $\Phi_0$, at 20mK and every 100mK starting from 0.10K up to 1.30K. The solid red line connects the data points taken at 20mK.

**Fig. 2** Traces of resistance versus temperature at several values of magnetic flux for Al-1. Filled and open circles correspond to resistances taken at integer and half-integer flux quanta, respectively. Whereas sharp transitions to zero resistance were observed at integer $\Phi_0$, a broad drop characterized the behavior at $\frac{1}{2}\Phi_0$, where the resistance leveled off to a significant fraction of the normal state resistance at temperatures below 200mK. Lines are used to connect the data points.

**Fig. 3** Traces of resistance versus temperature at several values of magnetic flux for AuIn-1, a $Au_{0.7}In_{0.3}$ cylinder with $d$ = 154 nm and $t$ = 30 nm. Filled and open circles correspond to resistances taken at integer and half-integer flux quanta, respectively. The resistance at $\frac{1}{2}\Phi_0$ leveled off at about 0.80$R_N$, showing almost no change from 200mK down to 20mK. Lines are used to connect the data points. Inset: $R(\Phi)$ for AuIn-1 at the lowest temperature, $T$ = 20 mK. For most fields below $H_{c//}$ the sample was superconducting, except around $\Phi = \pm\frac{1}{2}\Phi_0$ where sharp

resistance peaks of width $\delta\Phi = 0.1\,\Phi_0$ were found. From $H_{c//} = 2382$ G ($\Phi_{c//} = 2.14\,\Phi_0$), $\xi$(20 mK) is estimated to be 160 nm.

**Fig. 4.** Traces of resistance versus temperature at several values of magnetic flux for Al-2, a larger Al cylinder with $d = 357$ nm and $t = 30$ nm. The superconducting coherence length $\xi$(T) is 60 nm at $T = 0.39$ K, as estimated from $H_{c//}$. Therefore at T = 0, $\xi(0) < 60$ nm, and $d > \xi(0)$. Insets: a) $R(\Phi)$ at several temperatures exhibited conventional Little-Parks resistance oscillations of period $\Phi_0 = h/2e$. b) Measured $\Phi$-$T$ phase diagram for Al-2. A single superconducting region, with a phase boundary modulated by an oscillation of period $\Phi_0 = h/2e$, was observed. A resistance value of $R = 400\,\Omega$ was used to determine the superconducting-normal phase boundary, $T_c(\Phi)$.

**Fig. 5** a) $\Phi$-$T$ phase diagram for Al-1 ($d = 150$ nm). Disconnected superconducting regions separated by a normal phase are found, even in the zero-temperature limit. The solid lines are fits to de Gennes' theory. A value of $R(T_c) = 0.05 R_N$ was chosen to determine the phase boundary, $T_c(\Phi)$. Note that the temperature range (0 - 1.5 K), is much larger than that shown for Al-2 (1.25 - 1.45 K) given in Fig. 4, Inset b.

**References and Notes**

1. A.K. Geim *et al.*, *Nature*, **396**, 144 (1998).
2. A.K. Geim, *et al.*, *Nature*, **390**, 259 (1997).
3. P. W. Anderson, *J. Phys. Chem. Solids* **11**, 26 (1959).
4. For a recent review, see M. Tinkham, D. C. Ralph, C. T. Black, J. M. Hergenrother, *Czech. J. Phys*. **46**, 3139 (1996).
5. P. -G. de Gennes, *C. R. Acad. Sc. Paris* **292**, 279 (1981).
6. V.V. Moshchalkov, *et al.*, *Nature*, **373**, 319 (1995).
7. B. S. Deaver Jr., W. M. Fairbank, *Phys. Rev. Lett*. **7**, 43 (1961).
8. R. Doll, M. Näbauer, *Phys. Rev. Lett*. **7**, 51 (1961).
9. C. N. Yang, *Rev. Mod. Phys*. **34**, 694 (1962).

10. P. -G. de Gennes, *Superconductivity of Metals and Alloys*. (W. A. Benjamin, Inc., New York, 1966), p. 185.

11. W. A. Little, R. D. Parks, *Phys. Rev. Lett.* **9**, 9 (1962).

12. For a recent review, see A. M. Goldman, N. Markovic, *Phys. Today* **51**, 39 (1998).

13. C. Kittel, *Introduction to Solid State Physics*, (John Wiley and Sons Inc., New York, ed. 7, 1996), p. 253.

14. Yu. Zadorozhny, D. R. Herman, Y. Liu, *Phys. Rev. B* **63**, 144521 (2001).

15. Yu. Zadorozhny, Y. Liu, *Europhys. Lett.* **55**, 712 (2001).

16. M. Tinkham, *Introduction to Superconductivity*, (McGraw Hill Inc., New York, ed. 2, 1996), p. 128.

17. A. Kapitulnik, N. Mason, S. A. Kivelson, S. Chakravarti, *Phys. Rev. B* **63,** 125322 (2001).

18. A. Yu. Kasumov, *et al.*, *Science* **284**, 1508 (1999).

19. See, for example, P. Singha Deo, V. A. Schweigert, F. M. Peeters, A. K. Geim, *Phys. Rev. Lett*. **79**, 4653 (1997); and references therein.

20. M. B. Sobnack, F. V. Kusmartsev, *Phys. Rev. Lett*. **86**, 716 (2001); and references therein.

21. We would like to thank A. M. Goldman and J. Jain for useful discussions, Sebastien Diaz for translating Ref. 5 into English, and K. D. Nelson, Z. Q. Mao, E. Hutchinson and D. Okuno for technical assistance. This work is supported by grants from the U.S. National Science Foundation and the Penn State MRSEC.

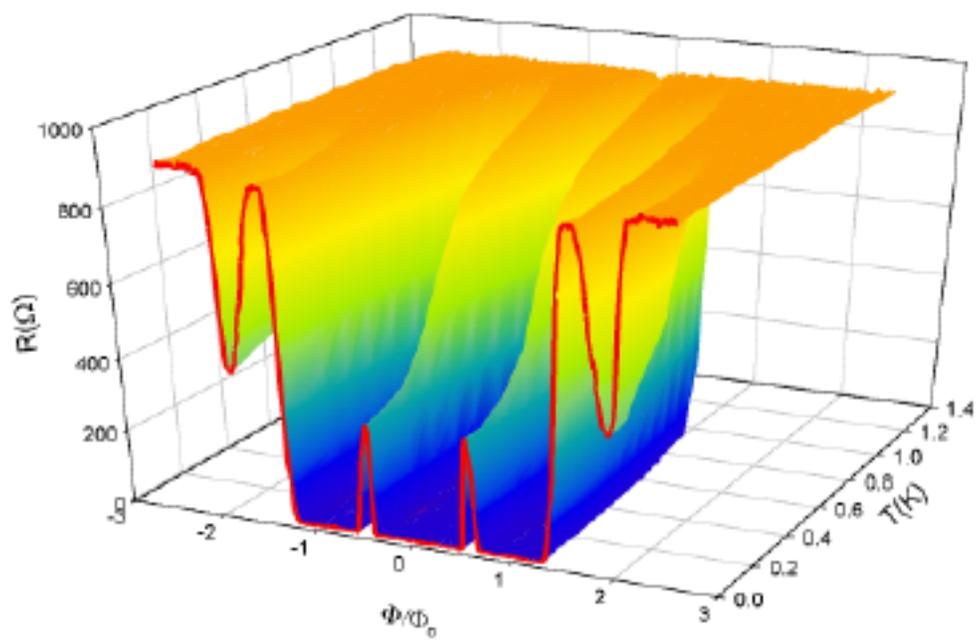

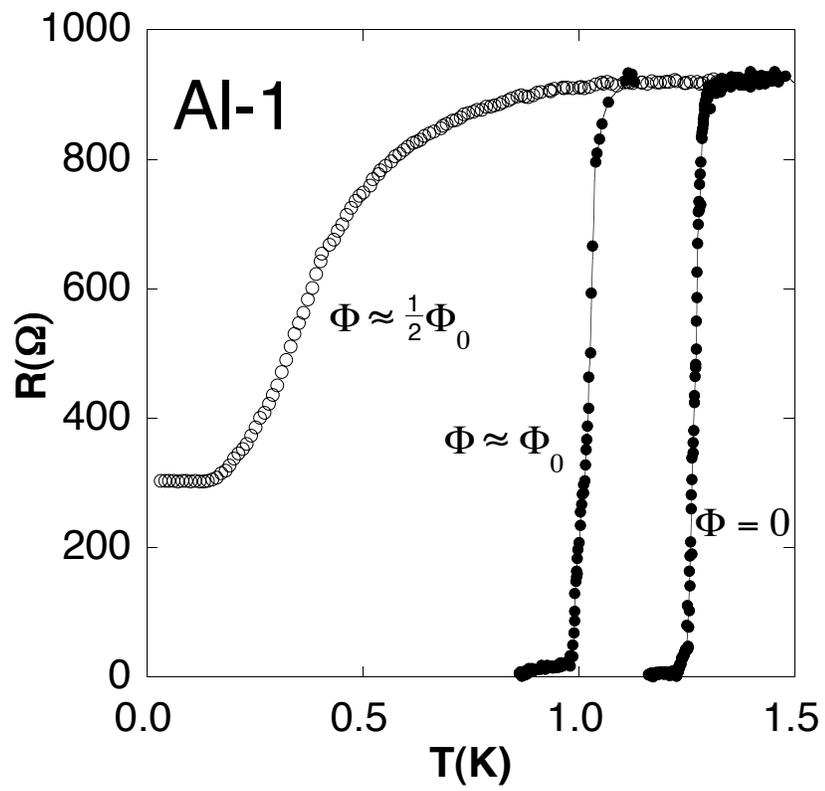

Figure 2

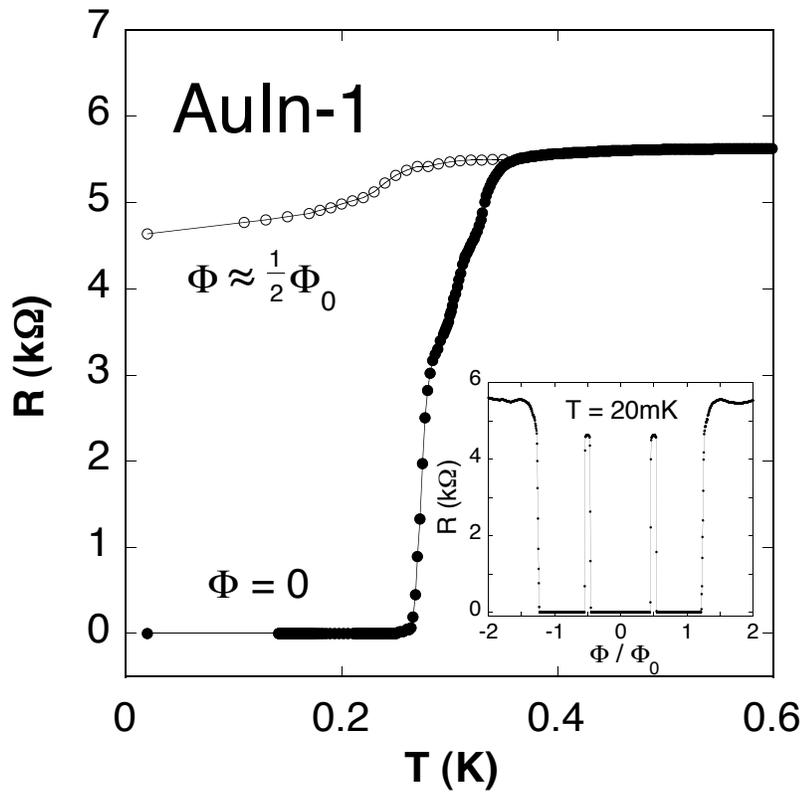

Figure 3

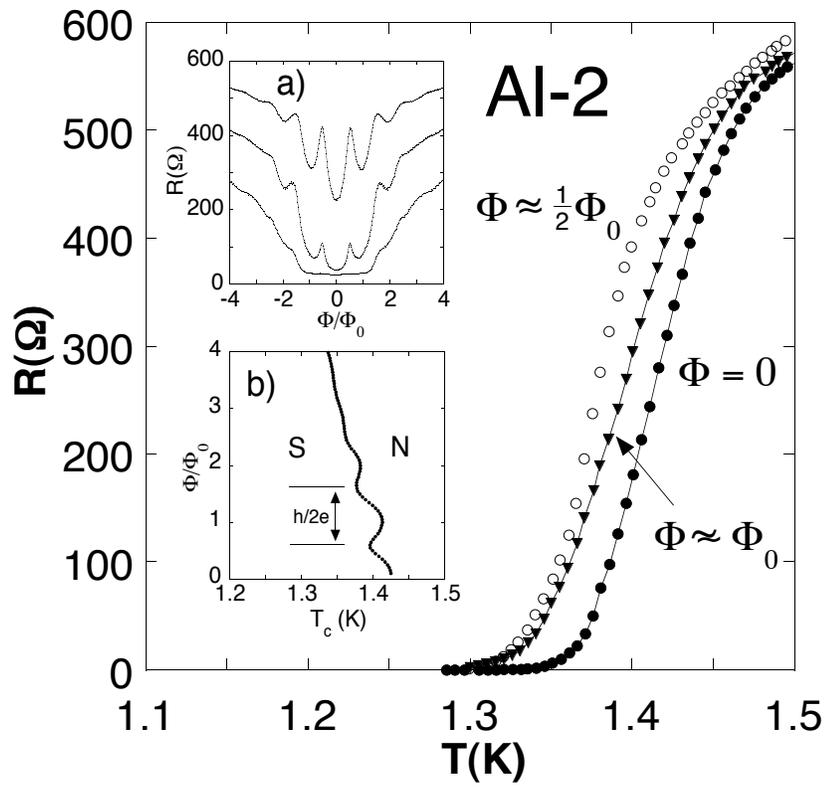

Figure 4

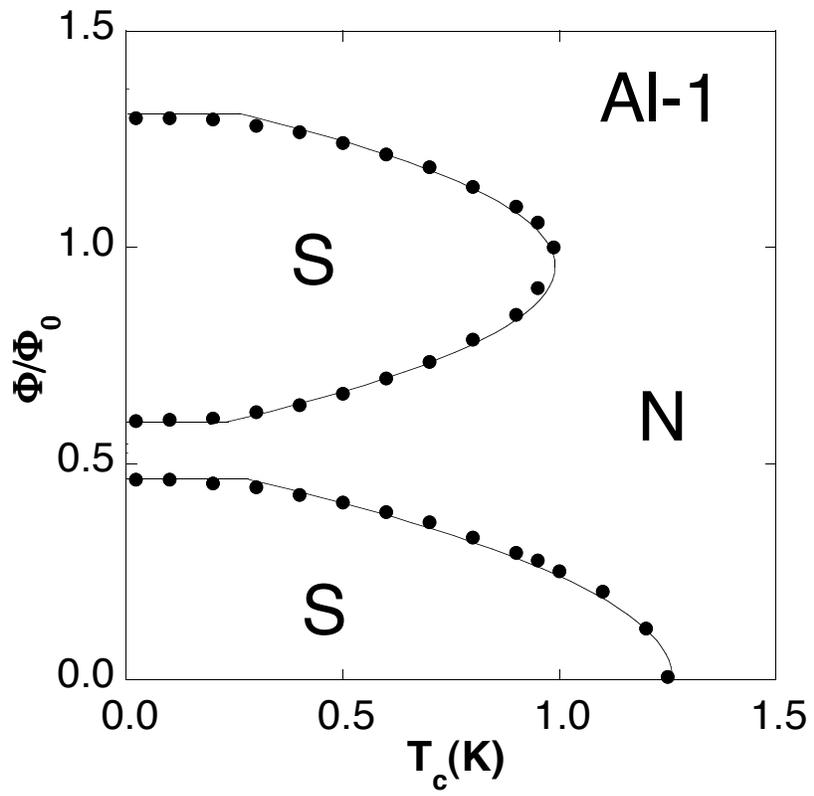

Figure 5